\newcommand{\dint}{{\rm d}}
    \documentclass[%
    reprint,
    amsmath,amssymb,
    aps,
]{revtex4-1}
    \usepackage{amsmath}
    \usepackage{cleveref}
    \usepackage{amsfonts}
    \usepackage{graphicx}
    \usepackage{dcolumn}
    \usepackage{bm}

    \usepackage{titlesec}
  \usepackage{easyReview}
    \titlespacing{\subsubsection}{2pt}{\parskip}{-\parskip}
    
\begin{document}

    	\title{ Energy dependence of the proton geometry in exclusive vector meson production}

    	\author{Arjun Kumar}
    	\email{arjun.kumar@physics.iitd.ac.in}

    	\author{Tobias Toll}%
    	\email{tobiastoll@iitd.ac.in}
    	\affiliation{%
    		Department of physics, Indian Institute of Technology Delhi, India\\
    	}

    	\date{\today}

    	\begin{abstract}
		The gluon radius of the proton is expected to increase at small gluon momentum fractions $x$, an effect which has hitherto not been considered in the dipole model framework. We investigate the energy dependence of exclusive $J/\psi$, $\phi$, and $\rho$ production by introducing three models for $x$ dependence of the gluon thickness function. We allow the transverse width of the  proton to increase as $x$ decreases, using novel parametrisations in the spherical proton and  the hotspot model. We compare these with a model where the number of hotspots increases as $x$ decreases and confront the models with HERA data. The models exhibit clear differences in the slope of the $t$-spectra and in the cross section ratios between coherent and incoherent events. Comparisons to $t$-slopes and $W_{\gamma p}$ measurements show a preference for models where the proton's size increases as $x$ decreases.  
    	\end{abstract}

    	\maketitle

    	\section{\label{introduction}Introduction}
    	Diffractive vector meson production is characterized by a momentum exchange with vacuum quantum numbers in the Mandelstam \emph{t}-channel between the virtual photon and the target in \emph{ep}  and \emph{e}A collisions \cite{Aktas:2005xu,Chekanov:2002xi,ZEUS:2007iet}. In the target rest frame, this is described by the dipole picture \cite{GolecBiernat:1998js, GolecBiernat:1999qd, Kowalski:2003hm, Kowalski:2006hc, Rezaeian:2012ji, Mantysaari:2018nng, Sambasivam:2019gdd} where the virtual photon splits up into a quark anti-quark dipole which interacts with the target via the strong interaction and then forms the final state. In a coherent event, the target stays intact, while in a dissociative or incoherent event, the target subsequently breaks into fragments. Coherent and incoherent diffractive events have been extensively studied at the HERA \emph{ep} experiments H1 and ZEUS \cite{Alexa:2013xxa,Chekanov:2009ab,H1:2009cml}. The coherent events are sensitive to the transverse structure of the target, while the measurements of incoherent events provide information about fluctuations in the target wave function.

	The measurements of inclusive deep inelastic scattering (DIS) at HERA and its theoretical description with the collinear factorisation framework with DGLAP evolution equations led to an understanding that gluons exist in the form of quantum fluctuations inside the proton, and their density increases steeply at higher energies. Though this framework has been very successful in dealing with hard processes, it could not explain diffractive events. The gluon density is also expected to saturate at small gluon momentum fractions $x$ so that unitarity of the events is upheld, even if a clear signal for saturation is yet to be seen in experiments. The dipole framework provides a unified description of inclusive and diffractive events, as well as exclusive diffraction, at small $x$. It also is a natural model for describing saturation physics.  
	
	In the dipole picture, the impact parameter is a Fourier conjugate to the momentum transfer at the target vertex, thus one can study the transverse structure of the target through cross sections differential to the Mandelstam $t$ variable, only experimentally available in exclusive diffractive events. Exclusive vector meson production in \emph{ep} collisions, therefore, serves as a good probe of the gluonic radius of the proton in the transverse plane and for their fluctuations. The dipole amplitude in these models are usually parametrisations fitted to HERA reduced cross section measurements.
	
	 In \cite{Kowalski:2003hm} the authors introduced an explicit dependence on the target geometry in the form of a gluon thickness profile in the transverse plane. This profile was taken to be independent of collision energies. This geometrical description was further enhanced in \cite{Mantysaari:2016jaz} by introducing three hotspots of gluon density inside the proton, which were allowed to fluctuate. They were hence able to provide a geometrical description of the incoherent cross sections at HERA. However, these hotspots were still independent of the photon-proton collision energies in the events. 
	 
	 It is expected that the transverse gluonic radius of the proton will evolve and increase for small Bjorken-\emph{x} as more low momentum gluons are radiated. The increase in the gluonic radius of the proton is compatible with Regge theory predictions, where the extracted coherent slope of the $t$-distribution increases logarithmically with increasing $W_{\gamma p}$ \cite{Collins:1977jy,Donnachie:1994zb}. This phenomenon is known as Gribov diffusion \cite{Gribov:1973jg}. The analysis in \cite{Albacete:2016pmp} also supports the increase in hotspot width in hotspot models. Here the authors conclude that the transverse diffusion or growth of the hotspots with increasing collision energy is the primary dynamical process underlying the onset of the hollowness effect in $pp$ interactions. Further, they ascribe the measured growth of the total $pp$ cross section to this mechanism. 
Recent studies on diffractive vector meson production with the description of the proton using JIMWLK evolved Wilson lines \cite{Mantysaari:2018zdd,Schlichting:2014ipa} also indicate the increase in width of  the fluctuating gluonic density hotspots with decreasing \emph{x}. Further, the investigations with light nuclei in \cite{Mantysaari:2019jhh} suggest an evolution of the transverse width of the fluctuating gluon distribution, which is incorporated in nuclei as sub-nucleon fluctuations. They also find that the JIMWLK evolution predicts an increased deuteron size at small $x$. Additionally, the fluctuations are expected to evolve in energy and at high energies, there should be no event-by-event fluctuations in the black-disc limit where dipole cross section saturates to unity. The incoherent cross section, sensitive to the event-by-event fluctuations, is vastly suppressed in this limit as shown in \cite{Mantysaari:2018zdd}. Recent data on $J/\psi$ photo-production in ultra-peripheral  collisions at low-\emph{x} from the ALICE collaboration at LHC \cite{ALICE:2018oyo} also indicate the suppressed incoherent cross section compared to coherent at high energies. Though the energy dependence of the incoherent cross section is still not  measured at ALICE, the $p_t$  distribution of the decay products of $J/\psi$ is indicative of the suppressed incoherent cross section. 
 
	 	 The dipole model provides a clean phenomenological environment to study the effects of energy evolution in proton geometry. It is, therefore, a fertile testing ground for implementing and comparing different aspects of energy dependence of the proton geometry and confront them with existing data as well as guide our understanding in what to expect from future colliders experiments such as the electron-ion collider (EIC) \cite{Accardi:2012qut,AbdulKhalek:2021gbh} and the large hadron-electron collider (LHeC) \cite{LHeC:2020van}. 

	In this study, we investigate the energy dependence of the initial state of the proton in exclusive $J/\psi$, $\rho$, and $\phi$ meson production using the saturated impact-parameter dependent dipole model bSat (also known as IP-Sat), and its linearised version, the bNonSat model.  We introduce a novel parametrisation of evolution effects in the proton geometry with and without subnucleon fluctuations. We do this by introducing an explicit $x$-dependence in the proton and hotspot widths. We also compare it with our implementation of the approach given in \cite{Cepila:2016uku}, in which the number of hotspots have an $x$-dependence. 
		 
        	 The paper is organised as follows. In the next section, we give an outline of exclusive vector meson production in the dipole formalism, as well as of the hotspot model. We then introduce our modifications to these models to take energy dependence into account. In Section \ref{sec:Results} we present the results and compare our predictions with the available HERA data. Finally, we summarise and discuss the main conclusions of the study.

\section{\label{dipole_picture} The Color Dipole Models}
\begin{figure}
	\centering
	\includegraphics[width=0.65\linewidth]{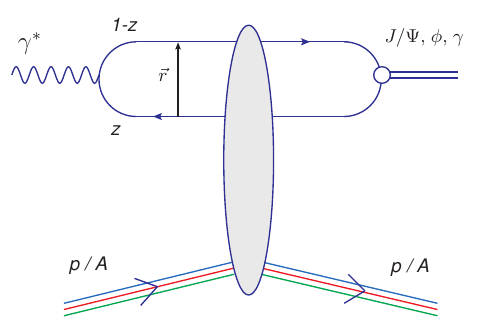}
	\caption{Exclusive vector meson production in the dipole picture of DIS. See description of all variables in the text. }
	\label{dipole}
\end{figure}

In the dipole picture, the scattering amplitude for the diffractive vector meson production factorises at high energy and is given by the convolution of three subprocesses, as depicted in Fig.~\ref{dipole}. First, the virtual photon splits into a quark anti-quark dipole; then, the dipole  interacts with the proton via one or many colourless two-gluon exchanges, after which it recombines into a vector meson. The amplitude is given by:
 \begin{align}
    	\mathcal{A}_{T,L}^{\gamma^* p \rightarrow J/\Psi p} (x_{I\!\!P},&Q^2,\textbf{$\Delta$})=i\int \dint^2 \textbf{r}\int \dint^2\textbf{b}\int \frac{\dint z}{4 \pi} \\ \nonumber &\times (\Psi^*\Psi_V)_{T,L}(Q^2,\textbf{r},z) \\ \nonumber 
	&\times e^{-i[\textbf{b}-(1-z)\textbf{r}].\textbf{$\Delta$}} \frac{\dint\sigma _{q\bar{q}}}{\dint^2\textbf{b}}(\textbf{b},\textbf{r},x_{I\!\!P})
 \end{align}
 where $L$ and $T$ represent the longitudinal and transverse polarisation of the virtual photon, \textbf{r} is the transverse size and direction of the dipole, $z$ is the energy fraction of the photon taken by the quark, \textbf{b} is the impact parameter of the dipole relative to the proton and $\bf{\Delta}$ is the pomeron's transverse momentum where $|{\bf \Delta}|=\sqrt{-t}$. Here, $(\Psi^* \Psi_V)$ is the wave-function overlap between the virtual photon and the vector meson in the final state. The exponential factor has two terms, the \textbf{b} term comes from the impact parameter space while the \textbf{r} term is due to non-forward wave functions which  calculated by Bartels, Golec-Biernat, and Peters in \cite{Bartels:2003yj}. This BGBK factor is important for studying $\rho$ and $\phi$ mesons at low $Q^2$. The amplitude is a Fourier transform from the transverse coordinate of the quark in the dipole to the transverse momentum transfer in the interaction. The amplitude thus contains information on the spatial structure and fluctuations of the gluon density inside the proton. The dipole cross section ${\rm d}\sigma _{q\bar{q}}/{\rm d}^2\textbf{b}$ describes the strong interaction. We use the boosted Gaussian wave function with the parameter values from \cite{Mantysaari:2018nng} for the vector-meson wave function. A recent update on different vector-meson wave functions can be found in \cite{Lappi:2020ufv}.
   
   The elastic diffractive cross section for a spherical proton (without geometrical fluctuations) is given by:
\begin{equation}
    	\frac{\dint \sigma^{\gamma^* p \rightarrow J/\Psi p}}{\dint t} = \frac{1}{16 \pi} \big| \mathcal{A}^{\gamma^* p \rightarrow J/\Psi p} \big|^2
\end{equation}
 We include initial state fluctuations in the proton's wave function by employing the Good-Walker formalism \cite{Good:1960ba}, where the coherent cross section is given by the first moment of the amplitude with respect to the initial state. In contrast, the total cross section is given by its second moment. The incoherent cross section thus probes the difference between the second and first moment squared, which for Gaussian distributions is the variance of the initial state. Thus, for an event-by-event variation $\Omega$, we have:
    \begin{eqnarray}
    	\frac{\dint \sigma_{\rm coherent}}{\dint t} &=& \frac{1}{16 \pi} \big| \left<\mathcal{A}(x_{I\!\!P},Q^2,\textbf{$\Delta$})\right>_\Omega\big|^2 \nonumber \\
    	\frac{\dint \sigma_{\rm incoherent}}{\dint t} &=& \frac{1}{16 \pi}\bigg(\big< \big| \mathcal{A}(x_{I\!\!P},Q^2,\textbf{$\Delta$})\big|^2\big>_\Omega \\\nonumber&~& ~~- \big| \big<\mathcal{A}(x_{I\!\!P},Q^2,\textbf{$\Delta$})\big>_\Omega\big|^2\bigg)
    	\end{eqnarray}
\begin{figure*}
	\centering
	\includegraphics[width=0.42\linewidth]{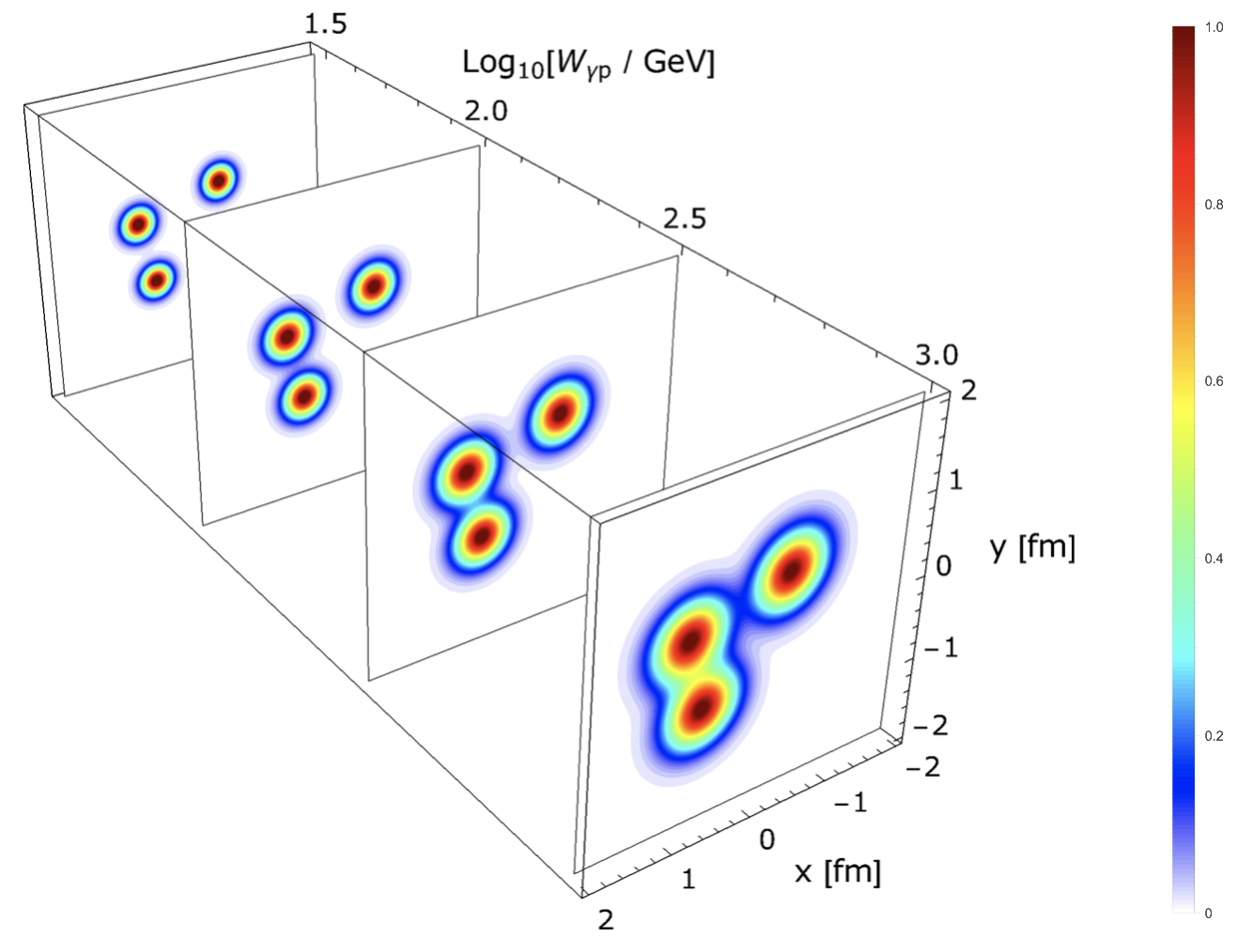}\hskip1.0cm
		\includegraphics[width=0.4\linewidth]{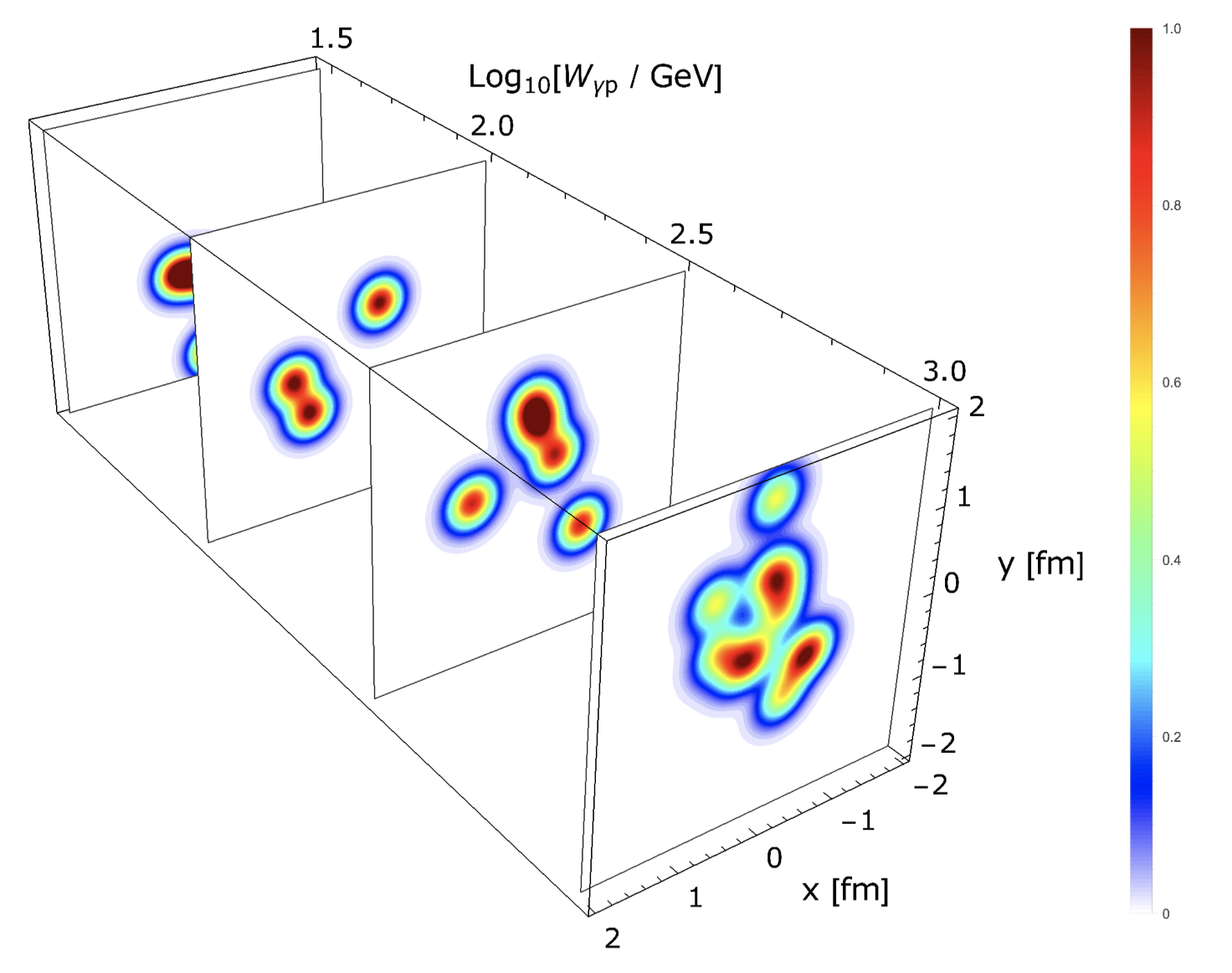}
	\caption{Transverse profile of an event of the proton with sub-nucleon fluctuations in the VHW model (left) and in the VHN model (right). The parameters are from the bSat model in table \ref{table}.}
	\label{profile}
\end{figure*}   

We consider two versions of the dipole cross section. The bSat model dipole cross section is given by \cite{Bartels:2002cj}:
 \begin{eqnarray}
    	\frac{\dint \sigma _{q\bar{q}}}{\dint^2\textbf{b}}(\textbf{b},\textbf{r},x_{I\!\!P})=
	2\big[1-\text{exp}\big(-F(x_{I\!\!P},\textbf{r}^2)T_p(\textbf{b})\big)\big],
\end{eqnarray}
with
\begin{eqnarray}
    	F(x_{I\!\!P} ,\textbf{r}^2)=\frac{\pi^2}{2N_C} \textbf{r}^2 \alpha_s(\mu^2) x_\mathbb{P} g(x_{I\!\!P},\mu^2).
\end{eqnarray}

Due to the exponential functional form in this case, the dipole cross section saturates for large gluon densities $xg(x, \mu^2)$  and for large dipole sizes $r$. 
The scale at which the strong coupling $\alpha_s$ and gluon density are evaluated is $\mu^2 = \mu_0^2 +\frac{C}{r^2}$ and the gluon density at the initial scale $\mu_0$ is parametrised as:
\begin{eqnarray*}
   x g(x,\mu_0^2)= A_g x^{-\lambda_g}(1-x)^{6}
\end{eqnarray*}
where the parameters $A_g, \lambda_g, C, m_f$ are determined through fits to inclusive reduced cross section measurements. 
 The bNonSat model is a linearised version of the bSat model where :
\begin{equation}
    	\frac{\dint \sigma _{q\bar{q}}}{\dint^2\textbf{b}}(\textbf{b},\textbf{r},x_{I\!\!P})=\frac{\pi^2}{N_C}\textbf{r}^2\alpha_s(\mu^2) x_{I\!\!P} g(x_{I\!\!P},\mu^2)  T_p(\textbf{b})
\end{equation}
which does not saturate for large gluon densities and large dipoles. This dipole cross section corresponds to a single two-gluon exchange. We use the fit results from \cite{Sambasivam:2019gdd}, which also includes a Gaussian suppression of large dipoles. 
The transverse profile of the proton is usually assumed to be Gaussian:
\begin{eqnarray}
T_p(\textbf{b}) = \frac{1}{2 \pi B_G}\exp\bigg(-\frac{\textbf{b}^2}{2B_G}\bigg)
\label{eq:profile}
\end{eqnarray}
and the parameter $B_G$ is constrained through a fit to the $t$-dependence of the exclusive J/$\psi$ production rates at HERA \cite{Kowalski:2006hc}, and is found to be $B_G=4\pm 0.4~$GeV$^{-2}$, which is fixed for all \emph{x} values. It should be noted that inclusive observables are not sensitive to the proton's profile, as this will only give contributions at higher twists. Therefore, as we modify the profile function in different ways in the rest of the paper it will not affect the quality of the fit for the bNonSat model at all, and only slightly for the bSat model. We have checked that the description of the HERA reduced cross section remains good in what follows. 

\subsection{The Fixed Hotspot Model}
Event-by-event fluctuations of initial state gluon density inside the proton can be taken into account by the hotspot model \cite{Mantysaari:2016ykx,Mantysaari:2016jaz,Mantysaari:2020axf}. Here, the gluons are assumed to be located in density hotspots. This can be implemented by changing the proton profile in eq.\eqref{eq:profile} in the following way:
\begin{eqnarray}
T_p(\textbf{b}) \rightarrow \frac{1}{N_{q}}\sum_{i=1}^{N_{q}}T_{q}(\textbf{b}-\textbf{b}_i),
\label{eq:hsprof}
\end{eqnarray}
with
\begin{equation}
T_{q}(\textbf{b}) = \frac{1}{2 \pi B_{q}} \frac{1}{\exp\big[\frac{\textbf{b}^2}{2B_{q}}\big]-S_g}
\label{eq:hsprofile}
\end{equation}
which is Gaussian for $S_{g} = 0 $ and peaks more at the centre for non-zero values of $S_g$ \cite{Kumar:2021zbn}. Here $N_{q}=3$ is the number of the hotspots located at $\textbf{b}_i$ sampled from a Gaussian width $B_{qc}$ and $B_{q}$ is the width of the hotspots.  $B_{qc}$ and $B_{q}$ control the amount of fluctuations in the proton geometry and are constrained by the coherent and incoherent data. For bNonSat, $S_{g} = 0$, while for bSat we use $S_g=0.3$ for $J/\psi$-production and $S_g=0.4$ for $\rho$- and $\phi$-production. We refer to \cite{Mantysaari:2022ffw} for a recent Bayesian analysis of the parameters of the hotspot model. Here in the hotspot model, the number and the size of the hotspots is fixed for all $x_{I\!\!P}$ values, thus resulting in a fixed width of the proton.  We call this the Fixed Hotspot (FH) model. 

We also include the fluctuations in the saturation scale in our investigations following \cite{Mantysaari:2016ykx}. The experimentally observed multiplicity distributions and the rapidity correlations in $pp$ collisions need these fluctuations in order to describe the data  \cite{McLerran:2015qxa,Bzdak:2015eii}, the saturation scale fluctuations are incorporated by letting the saturation scale of hotspots fluctuate independently as follows: 

\begin{align}
P(\text{ln } Q_S^2/\big<Q_S^2\big>)=\frac{1}{\sqrt{2 \pi \sigma^2}} \text{exp}\bigg[-\frac{\text{ln}^2 Q_S^2/\big<Q_S^2\big> }{2  \sigma^2}\bigg]
\end{align}
The saturation scale is $Q_S^2( x, b)\equiv 2/r_S^2$, where $r_S$ is defined by solving $1/2=F(x ,r_S^2)T_p(b)$ \cite{Kowalski:2003hm} in the dipole amplitude. We can implement these fluctuations by changing the normalisation of the profile function. We use $\sigma=0.4$ for $J/\psi$-production and $\sigma=0.6$ for $\rho$- and $\phi$-production. These fluctuations play an important role for small $|t|$ \cite{Mantysaari:2016jaz,Kumar:2021zbn}.

The differential cross section receives significant corrections (discussed in detail in \cite{Kowalski:2006hc, Mantysaari:2016ykx}). Firstly, the dipole amplitude is approximated to be purely imaginary. However, the real part of the amplitude is taken into account  by multiplying the crosssection by a factor of (1+$\beta^2$) with $\beta = \tan\big(\lambda \pi/{2})$, and $\lambda = \partial \log (\mathcal{A}_{T,L}^{\gamma^*p\rightarrow Vp})/\partial \log(1/x_{I\!\!P})$. Secondly, to take into account that the two gluons may have different momentum fractions, a skewedness correction to the amplitude is introduced \cite{Shuvaev:1999ce}, by a factor $R_g(\lambda)= 2^{2 \lambda +3}/\sqrt{\pi} \cdot\Gamma (\lambda_g + 5/2)/\Gamma (\lambda_g+4)$ with $\lambda_g= \partial \log (x_{I\!\!P}g(x_{I\!\!P}))/\partial \log(1/x_{I\!\!P})$.
In our model, we calculate both these corrections using a spherical proton. They have significant contribution at low momentum transfer $|t|$ (around 40-60 \%).

Next we consider three separate modifications to the proton's gluon density distribution.

 \subsection{ Evolution effects on the proton's transverse size}
First, we explore these evolution effects in a spherical proton (without initial-state fluctuations). We introduce the evolution effects on the size of the proton by parametrising the width of the proton as a function of $x_{I\!\!P}$. This is implemented by introducing an  $x_{I\!\!P}$ dependence in the profile function by changing eq.\eqref{eq:profile} as follows: 
\begin{eqnarray}
   T_p(\textbf{b})  \rightarrow  T_p(\textbf{b}, x_{I\!\!P}),
  ~B_G(x_{I\!\!P})=B_{p} ~x_{I\!\!P}^{~\lambda_p}, 
\end{eqnarray}
where $B_{p}$ determines the normalisation and $\lambda_p$ governs the evolution of the width of the proton in the transverse plane. As $x_{I\!\!P}=(M_V^2+Q^2+|t|)/(W_{\gamma p}^2+Q^2-m_p^2)$  (where $M_V$ and $m_p$ are the vector-meson and proton masses, respectively), the parameter values are determined through a fit to the $t$-spectra at different  $W_{\gamma p}$ values of exclusive $J/\psi$ production rates at HERA. The gluonic radius for the spherical proton is $r_{\rm rms}= \sqrt{2 B_G(x_{I\!\!P})}$ which now varies with Bjorken-$x$. 

 \subsection{ \label{fluctuations} Evolution effects on the hotspots' transverse size}

\begin{figure*}
	\centering
	\includegraphics[width=0.42\linewidth]{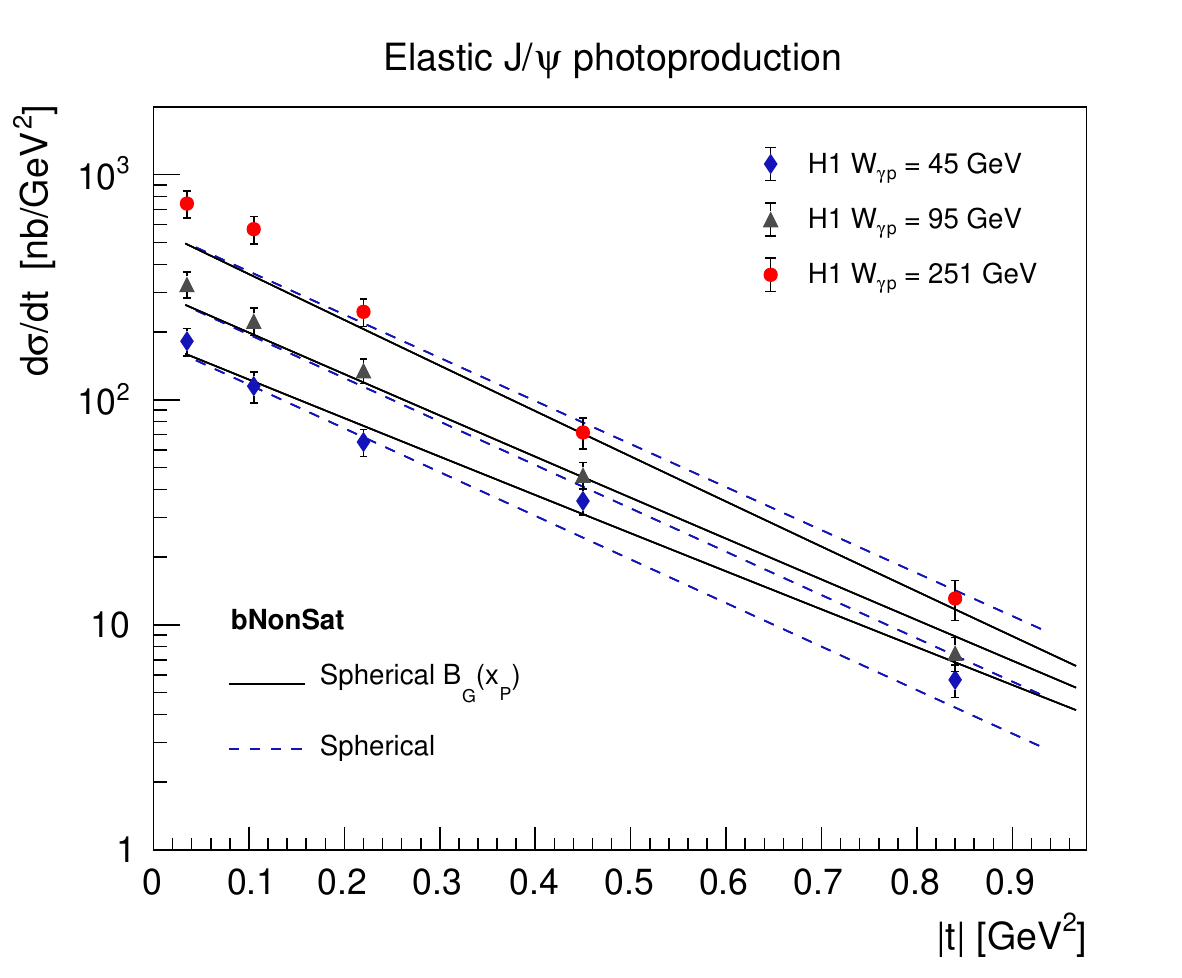}\hskip0.99cm
	\includegraphics[width=0.42\linewidth]{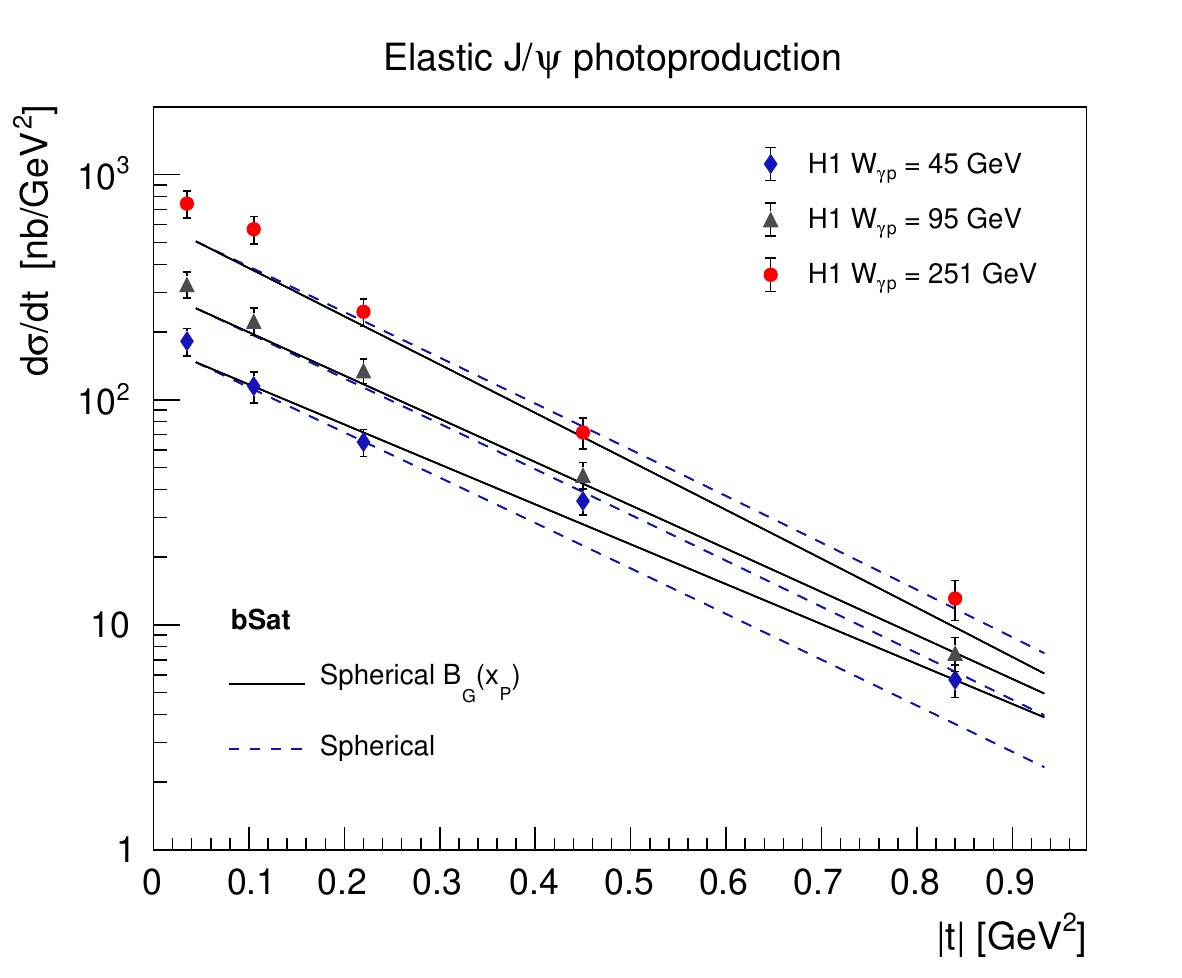}
	\includegraphics[width=0.45\linewidth]{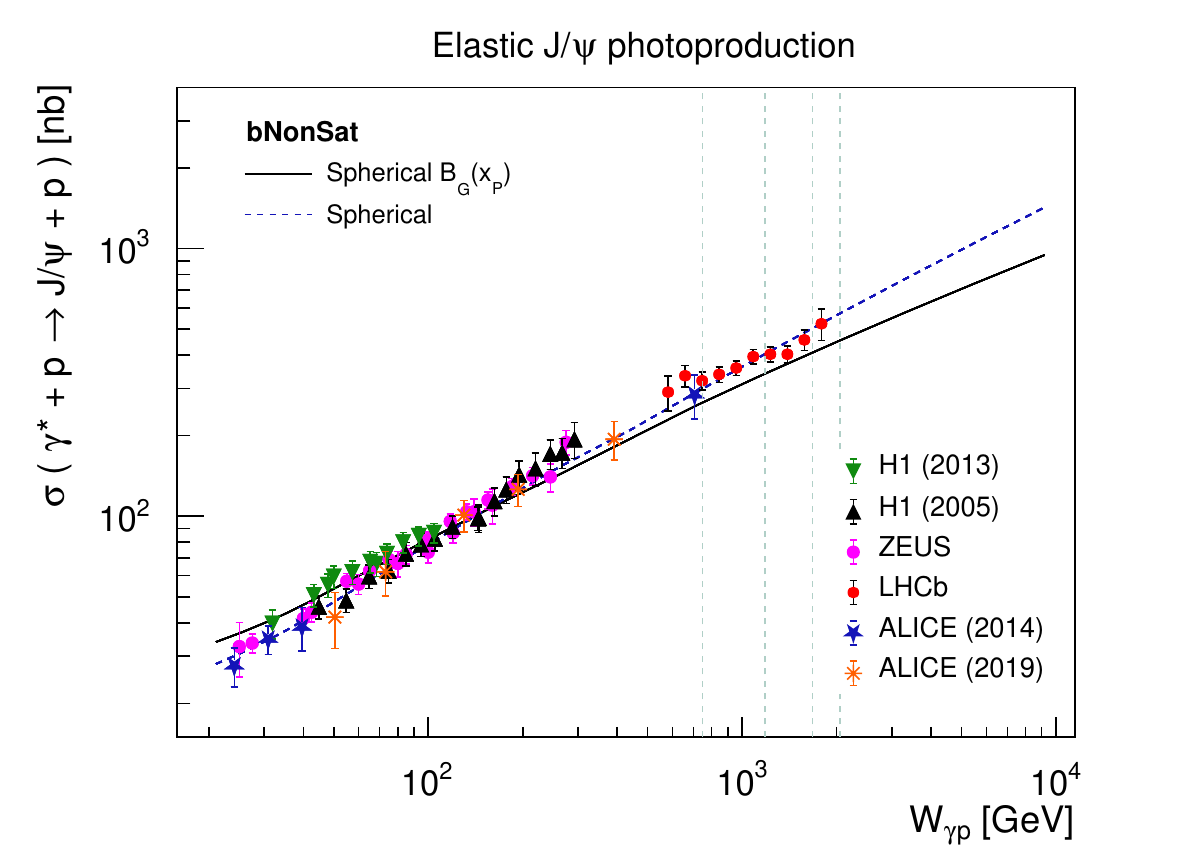}\hskip0.5cm
	\includegraphics[width=0.45\linewidth]{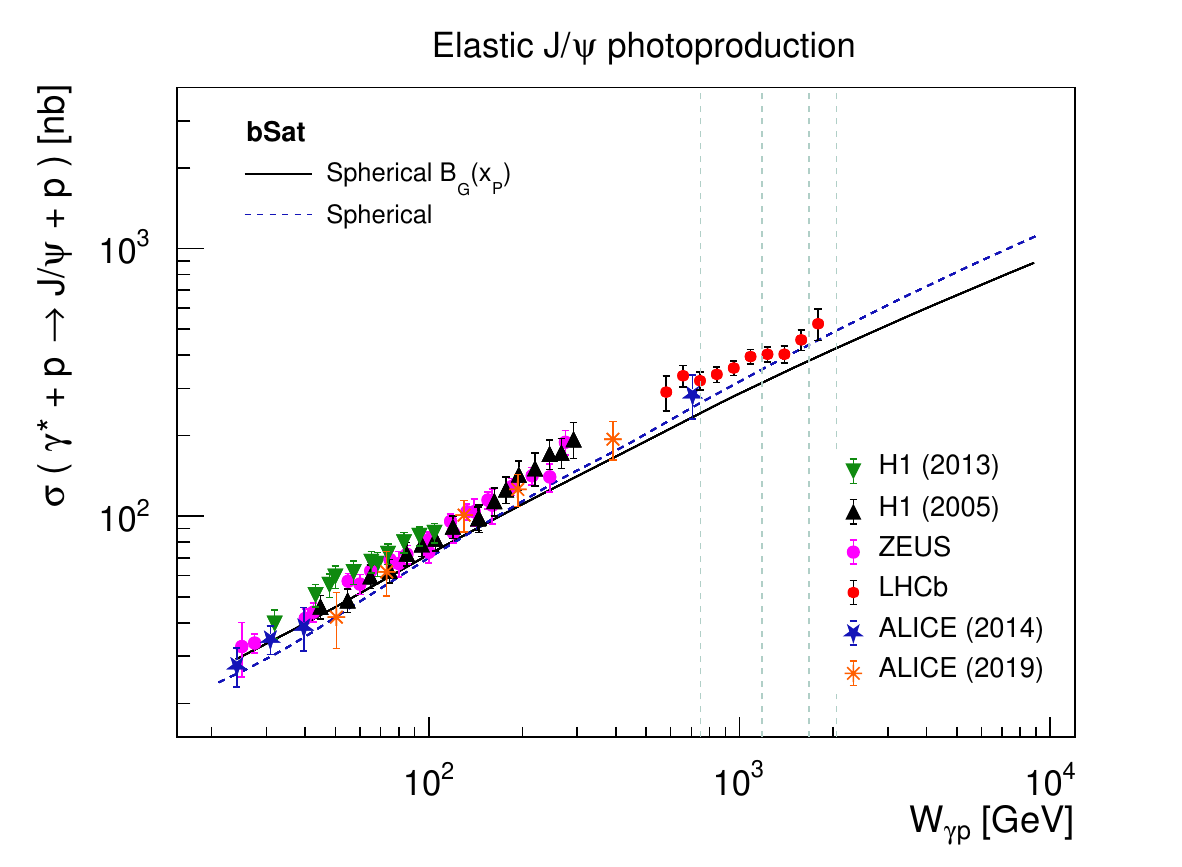}
	\caption{$t$-dependence (first row) and the energy dependence (second row) of the exclusive J/$\psi$ photo-production for a spherical proton (without fluctuations) in the bNonSat (left) and bSat (right) models with and without evolution effects. The experimental data is taken from  \cite{Alexa:2013xxa,Aktas:2005xu,Chekanov:2002xi,LHCb:2013nqs,ALICE:2014eof,ALICE:2019tqa}.}
	\label{tdep_Wdep_jPsi}
\end{figure*}  
To include the evolution effects of the proton's size in the hotspot model we make the width of hotspots $x_{I\!\!P}$-dependent in two different ways. As $x_{I\!\!P}$ decreases we expect more emissions of small $x_{I\!\!P}$ gluons inside the hotspots, thus increasing the hotspot size in the impact parameter space. We first implement this by changing the profile function in eq.\eqref{eq:hsprof} as follows:
\begin{eqnarray}
T_p(\textbf{b})\rightarrow \frac{1}{N_{q}}\sum_{i=1}^{N_{q}}T_{q}(\textbf{b}-\textbf{b}_i, x_{I\!\!P})
\end{eqnarray}
having the same profile of hotspots as in eq.\eqref{eq:hsprofile} with, 
\begin{equation}
B_{q} \rightarrow B_{q}(x_{I\!\!P})=B_{hs} ~x_{I\!\!P}^{~\lambda_{hs}} 
\end{equation}
where $B_{hs}$ determines the normalisation and $\lambda_{hs}$ governs the evolution of the width of hotspots and hence the width of the proton in the transverse plane.  We call this model the Varying Hotspot Width (VHW) model. 

In the models where the nucleon size grows, for instance \cite{Schlichting:2014ipa,Mantysaari:2018zdd,Schenke:2022mjv}, the Froissart bound \cite{Froissart:1961ux,Martin:1962rt} limits the energy dependence of the radius of hadron to be logarithmic or weaker. Hence we also consider a parametrisation for width of hotspots as 
\begin{eqnarray}
	B_q (x_{I\!\!P}) = b_0\ln^2\left(\frac{x_0}{x_{I\!\!P}}\right)
\end{eqnarray}
which results in a logarithmic increase of the radius of the proton. We will refer to this parametrisation as the 'logarithmic model'. A similar parametrisation with a quadratic dependence on the rapidity was considered to implement the growth of hotspots in a recent analysis \cite{Salazar:2021mpv} within the color glass condensate framework \cite{Gelis:2010nm,Iancu:2003xm} to study the $J/\psi$ production in  $pp$ and  $p$Pb collisions. We compare both the logarithmic and power law parametrisations for the  growth of the hotspots' width to show that the power law growth with small values of the evolution parameter  $\lambda_{hs}$  also follow the unitarity principle in the kinematic region considered in this study.

The gluonic radius for the proton in these cases is given by  $r_{\rm rms}= \sqrt{2(B_{qc}+B_{q}(x_{I\!\!P}))}$. Thus the average gluonic radius of the proton also increases with decreasing \emph{x} in addition to the increase in size of the hotspots. At large energies, as the size of the hotspots increase, the hotspots will begin to overlap, which will result in less fluctuations in the proton profile. Hence, at large energies, we expect a suppression of the incoherent cross section in these models.

The transverse profile of the proton with a fluctuating gluon distribution as a function of the centre of mass energy for the photon-proton system $W_{\gamma p}$ in VHW model is illustrated in Fig.~\ref{profile} (left).

 \subsection{  Evolution effects on the number of hotspots}
We next investigate another $x_{I\!\!P}$-dependence of the hotspot model, for which the number of hotspots increases with decreasing $x_{I\!\!P}$, while keeping the hotspots' width fixed. We refer to this model as the Varying Hotspot Number (VHN) model. This kind of a model was introduced in \cite{Cepila:2016uku}. In that analysis the authors considered the GBW parametrisation of the dipole amplitude and assumed a factorised form for the thickness function. Following \cite{Cepila:2016uku}, the number of hotspots in the model is parametrised as a function of $x_{I\!\!P}$. In this model the profile is the same as given in eq.(\ref{eq:hsprofile}) with the number of hotspots increasing stepwise at small $x_{I\!\!P}$ as:
 \begin{align}
 N_{q} \rightarrow N_{q}(x_{I\!\!P})=p_0~x_{I\!\!P}^{~p_1}(1+p_2\sqrt{x_{I\!\!P}})
 \end{align}
 We implemented this prescription into the dipole models described above. The hypothesis underlying this implementation is that as Bjorken-\emph{x} decreases, the gluon density increases steeply and instead of gluon accumulating in the original three hotspots of the FH model, the number of hotspots itself grows. The parameters $p_0$, $p_1$ and $p_2$ are determined through a simultaneous fit to the differential and the total incoherent cross sections for $J/\psi$ production. The VHN model is expected to exhibit less fluctuations at high energy as the numerous hotspots begin to overlap, which hence suppresses the incoherent cross section. The profile of the proton in the VHN model in illustrated in Fig.~\ref{profile} (right). 
 
\section{\label{sec:Results} Results}
\begin{table*}

	\begin{tabular}{lcccc}
			\hline
	\textbf{FH} &  VM~& $B_{q}$ (GeV$^{-2}$)~  &~$\chi^2/{\rm ndf}$  \\
		\hline
			bSat~~&$J/\psi$ &~0.985(50) ~ &~71.95/55 \\
		bNonSat~~& $J/\psi$&~0.935(30) ~ &~66.76/55\\
bSat~~&	$\rho, \phi$		 &~2.0(5) ~ &~301.84/66 \\
		\hline
\end{tabular}\hskip1.25cm
\begin{tabular}{lcccc}
	\hline
	\textbf{VHW}&  VM~& $B_{hs}$~(GeV$^{-2}$)~  &  $\lambda_{hs}$~&~$\chi^2/{\rm ndf}$  \\

	\hline
	bSat~~&$J/\psi$ &~0.245(10) ~&-0.213(7)~ &~72.23/54 \\
	bNonSat~~&		$J/\psi$&~0.256(9) ~&-0.1980(45)~ &~83.87/54 \\
	\hline
\end{tabular}\vskip0.25cm

	\begin{tabular}{lcccc}
	\hline
\textbf{Logarithmic}&	VM~& $b_{0}$~(GeV$^{-2}$)~  &  $x_{0}$~&~$\chi^2/{\rm ndf}$  \\
	\hline
	bSat/bNonSat&	$J/\psi$ &~0.075(4) ~&6.7(1.2)~ &~72.23/54 \\
bSat/bNonSat&	$\rho, \phi$ &~0.117(4) ~&20(6)~ &~458.53/65 \\
	\hline
	\end{tabular}

	\caption{Parameter values in the different dipole models described in the text. Uncertainties in the last digit(s) are shown in parentheses.}
	\label{table}
\end{table*}

\begin{figure*}
	\centering
	\includegraphics[width=0.42\linewidth]{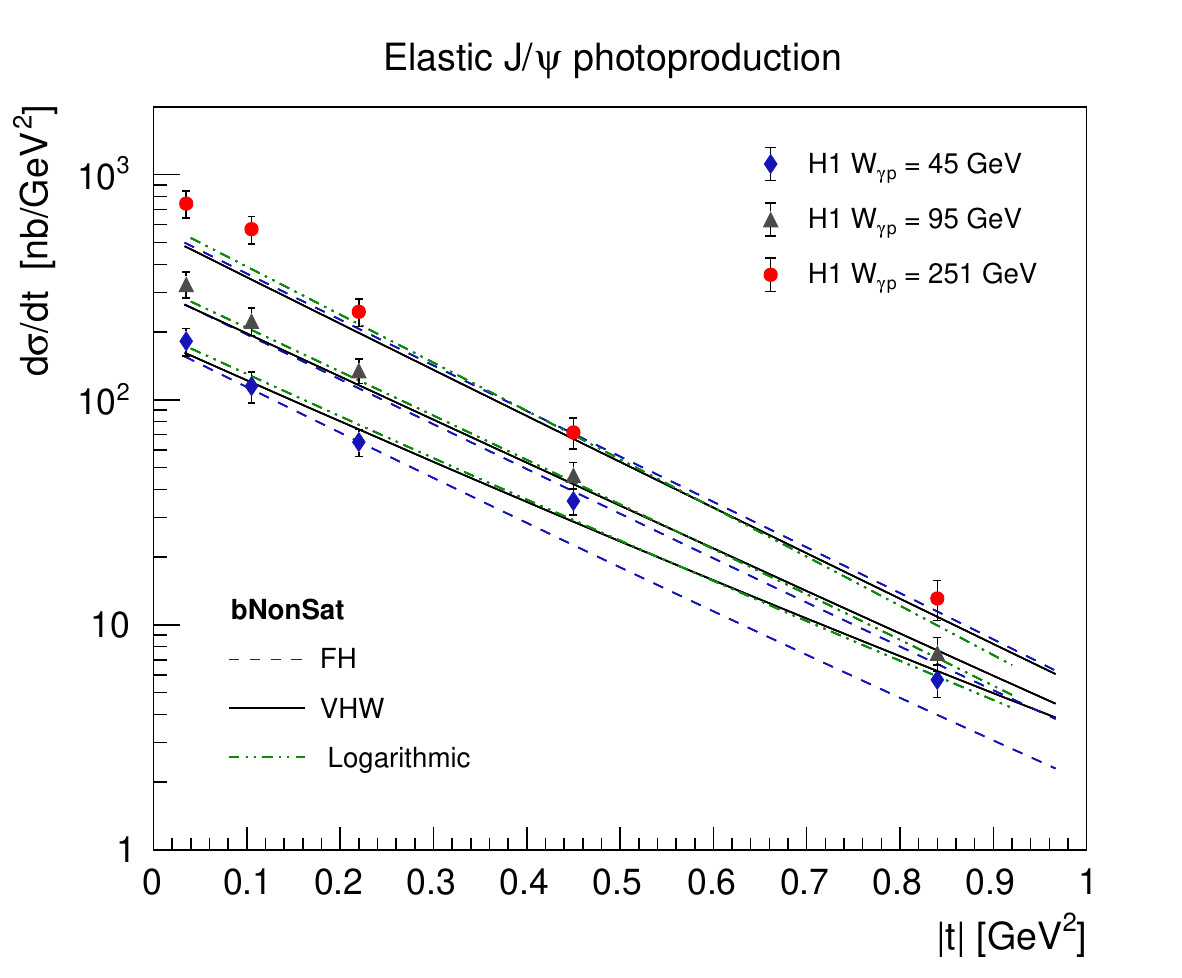}\hskip0.75cm
	\includegraphics[width=0.42\linewidth]{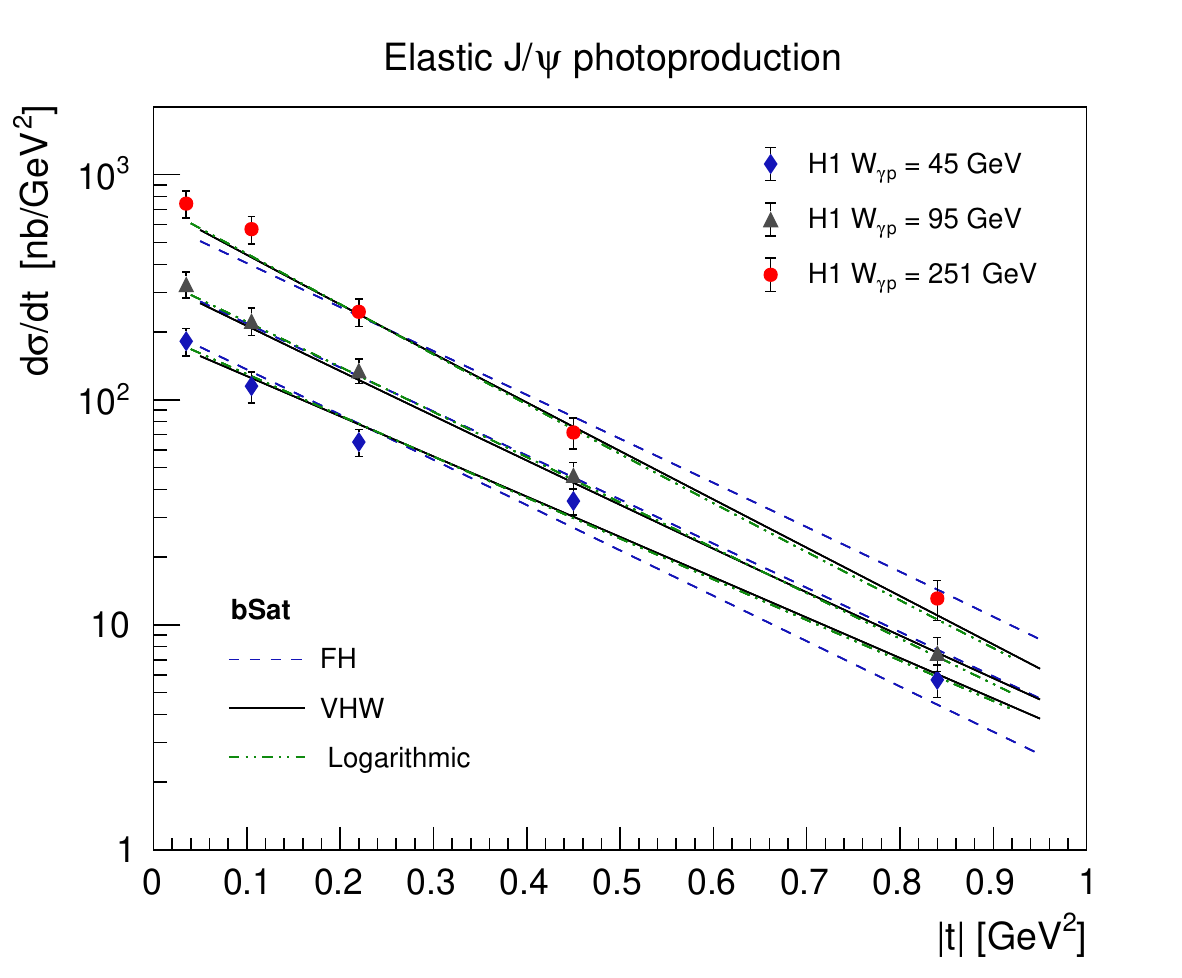}
	\caption{ $t$-dependence of the differential cross section for exclusive J/$\psi$ photo-production in the non-saturated (left) and saturated (right) FH, VHW, and logarithmic models fitted with the data from \cite{Aktas:2005xu}.}
	\label{tdep_BD_jPsi}
\end{figure*}  
 For a spherical proton with increasing transverse width, the parameters in the thickness function are determined through a fit to the $t$-dependence of exclusive $J/\psi$ cross sections at HERA at different $W_{\gamma p}$ \cite{Aktas:2005xu}. The optimal values of the parameters are found to be  $B_{p} = 2.3 \pm 0.03~ $GeV$^{-2}$ and $\lambda_p=-0.062 \pm 0.002$  with a $\chi^2/{\rm ndf} =  48.06/48$. The transverse gluonic radius of the  proton now increases at small $x_{I\!\!P}$ for these parameter values. 

In table \ref{table}, we provide the values of all the parameters in the FH and VHW models with and without saturation, as well as for the logarithmic model. The parameters values are determined through a simultaneous fit to the $t$-dependence of coherent and incoherent vector meson HERA measurements.  For J/$\psi$ photo production, the fit is done to the available $t$-dependence data \cite{Aktas:2005xu, Alexa:2013xxa} in different bins of W in the kinematic region $45 \leq W \leq  251$  GeV and $0.1 \leq t \leq 1.0$~GeV$^2$ while for the $\rho$ electro  production the fit is done to the available data  \cite{H1:2009cml, ZEUS:2007iet} in bins of $Q^2$ with  $0.1 \leq t \leq 1.0$~GeV$^2$. Note that for rest of the plots  the models  prediction are compared with the data. For the bSat case we use a modified profile originally introduced in \cite{Kumar:2021zbn} to explain the coherent and incoherent data well at small momentum transfer $|t|$. The poor description of the $\rho$ and $\phi$ data is partly due to tensions between the H1 and ZEUS measurements which makes a good fit to all available data difficult.
We implement the VHN model with parameters $p_0=0.011$, $p_1=-0.56$ and $p_2=165$ keeping the width of the hotspots the same as in the FH model in table \ref{table}. 
\begin{figure*}
	\centering
	\includegraphics[width=0.477\linewidth]{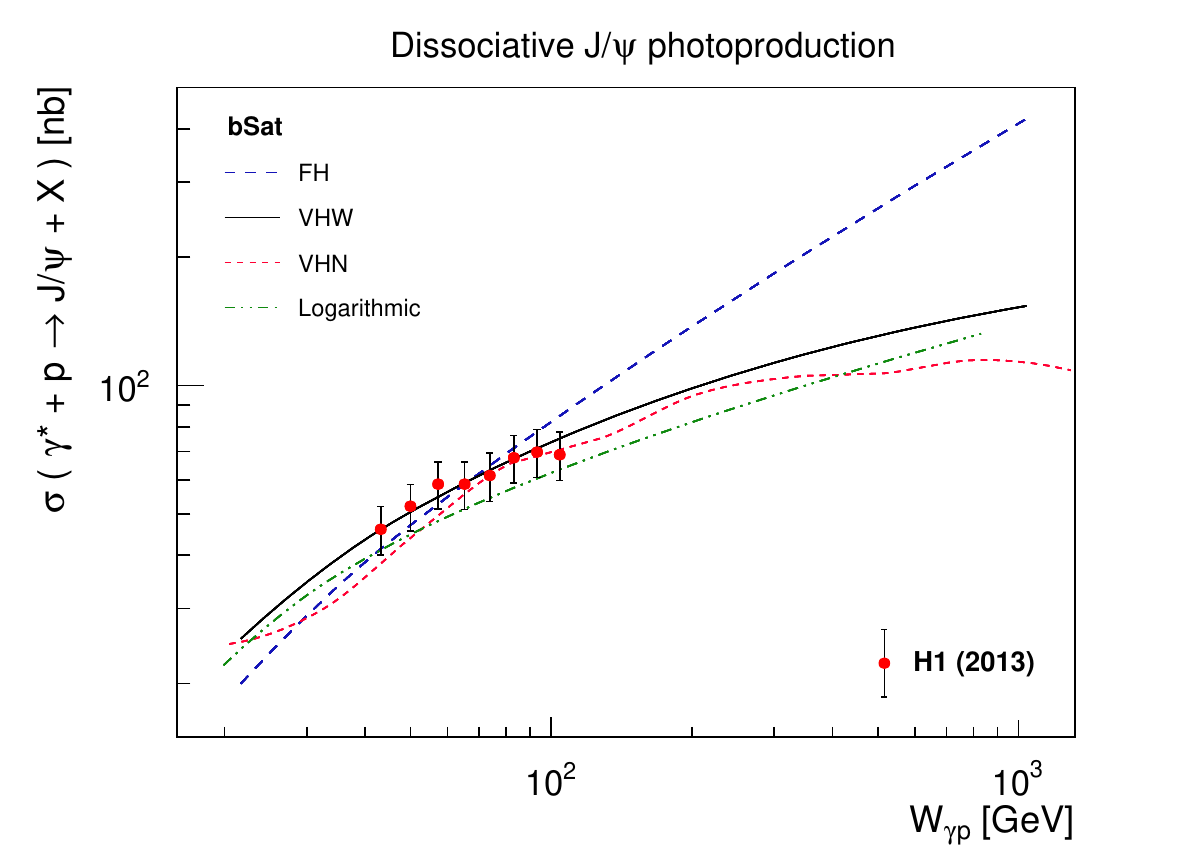}\hskip0.7cm
	\includegraphics[width=0.477\linewidth]{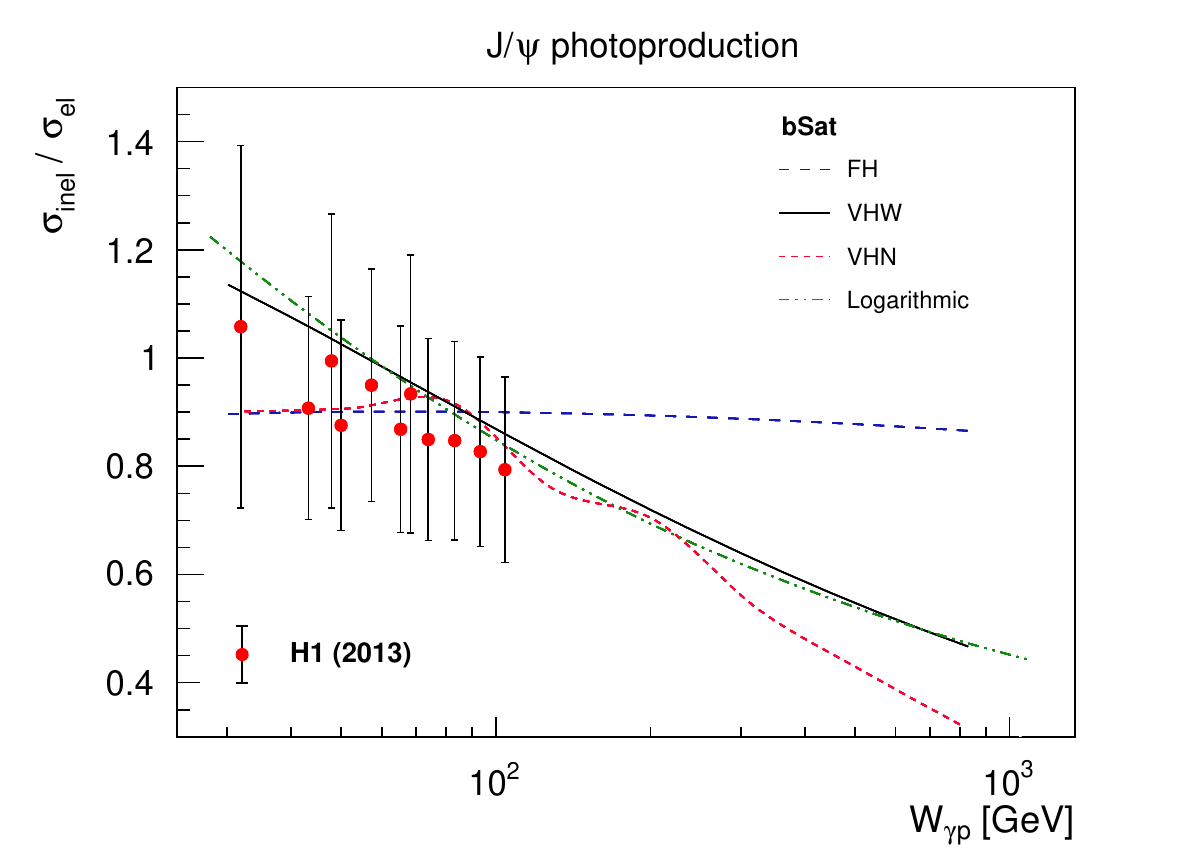}
	\caption{Energy dependence of the incoherent cross section (left) and the ratio of incoherent to coherent cross section (right) for $J/\psi$ photo-production in the saturated FH, VHW, VHN, and logarithmic models compared with the data from \cite{Alexa:2013xxa}.}
	\label{Wdep_R_jPsi}
\end{figure*}  

Fig.~\ref{tdep_Wdep_jPsi} depicts the $t$-dependence (first row) and the energy dependence (second row) of the exclusive J/$\psi$ photo-production for a spherical proton, with and without evolution effects for different values of $W_{\gamma p}$. 
We  obtain a similarly good agreement with the current data for the energy dependence in both the models with and without evolution effects. In the models with evolution effects, the total cross section increases at small $W_{\gamma p}$ and is suppressed at large $W_{\gamma p}$. For the bSat case, the model with evolution effects seems to underestimate the LHCb data points. 

In Fig.~\ref{tdep_BD_jPsi} we show the $t$-spectrum of the exclusive J/$\psi$ photo-production in the saturated and non-saturated FH and VHW models as well as the logarithmic model. The $t$-spectrum description in the VHW and the logarithmic models are very similar. As the hotspot width increases, it leads to an increase in the average gluonic radius of the proton and as a result the slope of the $t$-spectrum increases at higher energies. We have not shown the VHN model predictions as they are identical to that of the FH model. 
We see that even though the shapes of the $W_{\gamma p}$ and $t$ spectra change, the overall description of the measurements remains equally good. We see that for small $W_{\gamma p}$, the $t$-slope decreases while for large $W_{\gamma p}$ it increases, and the measured data seems to prefer a smaller slope overall. These effects off-set each other in the quality of the fits of the models. 
We further note that the bNonSat and bSat models give very similar predictions for the evolution effects on the size of the proton. 
This is consistent with earlier studies \cite{Mantysaari:2018nng,Kumar:2021zbn}, which have not been able to separate the two models for describing available $ep$ measurements. However, comparisons with UPC measurements with heavy ions at the LHC and RHIC experiments show a clear preference for the bSat model \cite{Sambasivam:2019gdd}. Hence, the rest of this study will be restricted to the bSat model.

In Fig.~\ref{Wdep_R_jPsi}, we study the energy dependence of the incoherent cross section as well as the ratio of incoherent to coherent cross sections for $J/\psi$ photo-production in the saturated FH, VHW  and VHN models as well as the logarithmic model. Many model-dependent effects get cancelled in such a ratio, such as real- and skewedness-corrections and other model uncertainties, making it a clear measure of the energy-dependence of suppression of initial state fluctuations in the proton. In the  VHW model we obtain a higher cross section at small $W_{\gamma p}$ while the cross section is suppressed at large energies. As the hotspot size increases in the VHW model and the logarithmic model at small $x_{I\!\!P}$, the hotspots begin to overlap and as a result the fluctuations are reduced at higher energies. The hotspot similarly begin to overlap at small $x_{I\!\!P}$ in the VHN model resulting in the suppressed incoherent cross section. We also plot the ratio of incoherent to coherent cross sections. In the VHW  and the logarithmic models, this ratio decreases with $x_{I\!\!P}$, while it remains nearly constant for the FH model. For the VHN model, this ration is first constant and then falls in discreet steps as the number of hotspots are constant, $N_{q} = 3$ for $20\leq W_{\gamma p}  \leq 120~$GeV, and then increases in integer steps, which results in the wobbly shape of the VHN curve in Fig.\ref{Wdep_R_jPsi}. The current HERA data does not distinguish between these scenarios but theoretically there is a clear difference at large energies. Unlike reference \cite{Cepila:2016uku}, the incoherent cross section does not fall after reaching a maximum in this kinematic range, which is similar to what was found in the MV model with explicit JIMWLK evolution in \cite{Mantysaari:2018zdd}. 
\begin{figure*}
	\centering
	\includegraphics[width=0.4445\linewidth]{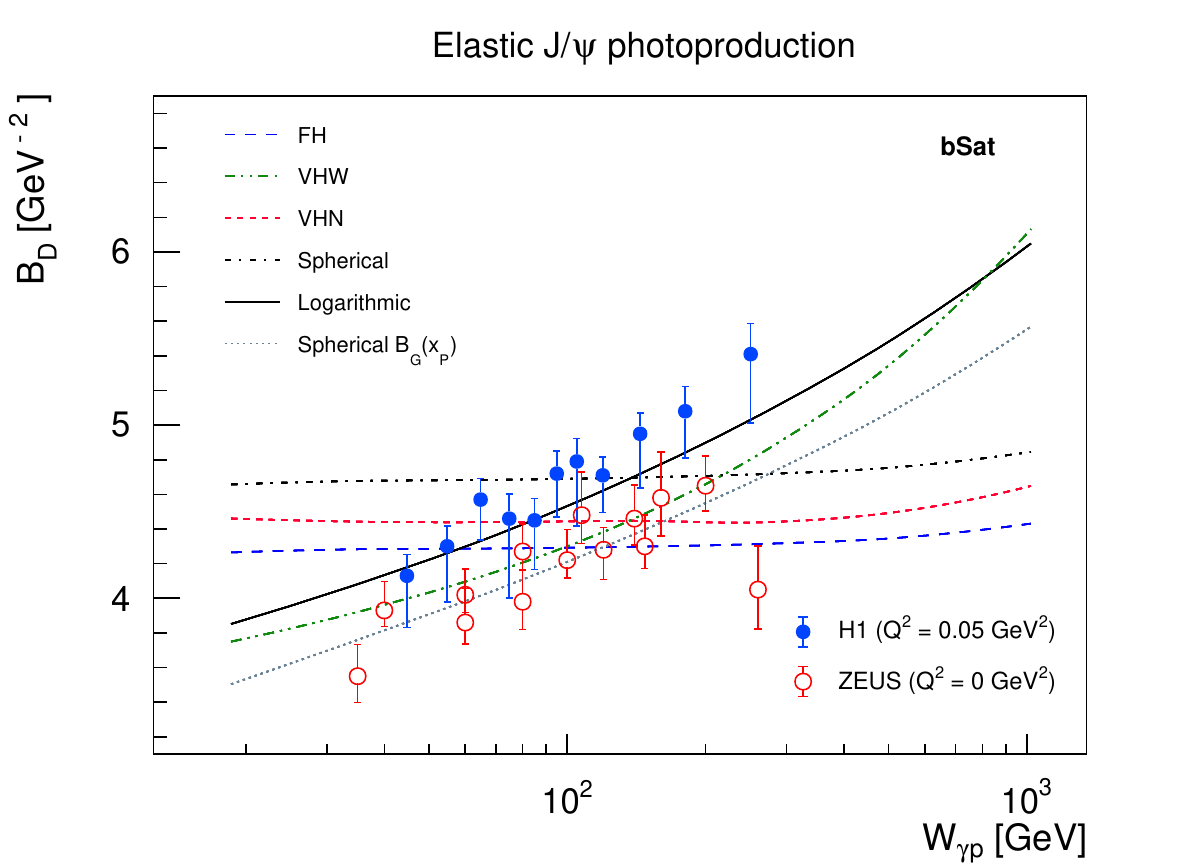} \hskip0.7cm
		\includegraphics[width=0.4445\linewidth]{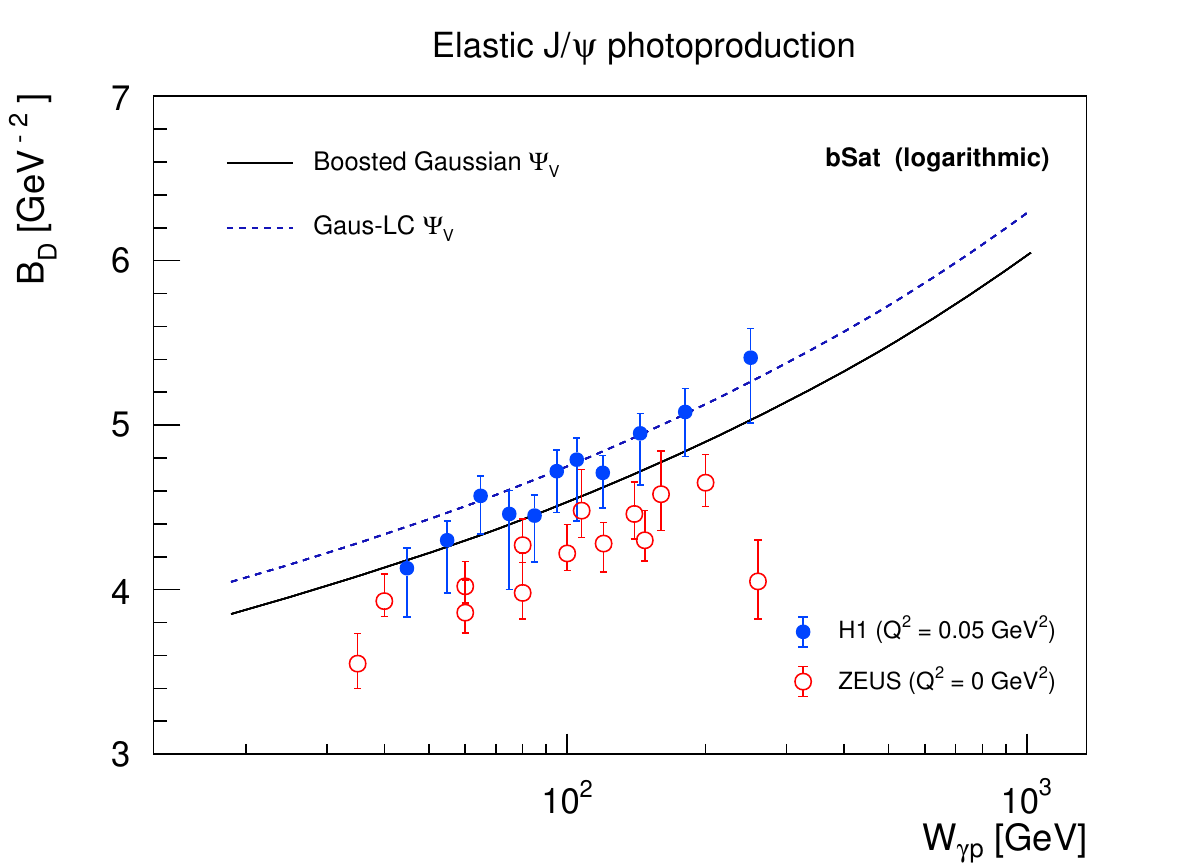}
	\caption{Left: Energy dependence of the extracted slope $B_D$ of the $t$-spectrum for exclusive J/$\psi$ photo-production in the  different models considered in this work. Right: A comparison of different vector meson wave function in the energy dependence of the extracted slope $B_D$ in the logarithmic model. The experimental data is from \cite{Aktas:2005xu,Chekanov:2002xi}. }
	\label{master}
\end{figure*}  

In the left panel of Fig.~\ref{master}, we plot the extracted coherent slope $B_D$ as a function of $W_{\gamma p}$ in all the models. Here, $B_D$ is defined by fitting  ${\rm d}\sigma/{\rm d}t\propto\exp(-B_D |t| )$ to the coherent $t$-distribution. This can be interpreted to correspond to the proton's effective transverse size. For the spherical proton with an energy-dependent width, and in the VHW model, the extracted slope clearly grows with $W_{\gamma p}$, faster for the latter. For the fixed sized spherical proton, as well as for the FH and VHN models, the transverse width remains constant even at higher energies. The logarithmic model represents the Froissart bound here and we see that all models lie below this for $W_{\gamma p} \lesssim 10^3$GeV which is where the power law of the VHW model crosses the $B_D$ value of the logarithmic model. The models' predictions are compared with the available HERA measurements which show a tendency that the average gluonic radius of the proton increases at higher energies. The H1 data shows a more pronounced increase in the size of the proton than the ZEUS data. In the right panel of Fig.~\ref{master} we test the dependence on the result on the choice of vector meson wave-function. We show the result from the logarithmic model using the default Boosted Gaussian wave-function as well as the Gauss-LC wave function \cite{Kowalski:2003hm}. We see that the uncertainty coming from the choice of wave-overlap is around 5\%, and that it does not affect the resulting shape of the $t$-slope with respect to $W_{\gamma p}$, which is consistent with \cite{Kowalski:2006hc}. 

\begin{figure*}
	\centering
	\includegraphics[width=0.45\linewidth]{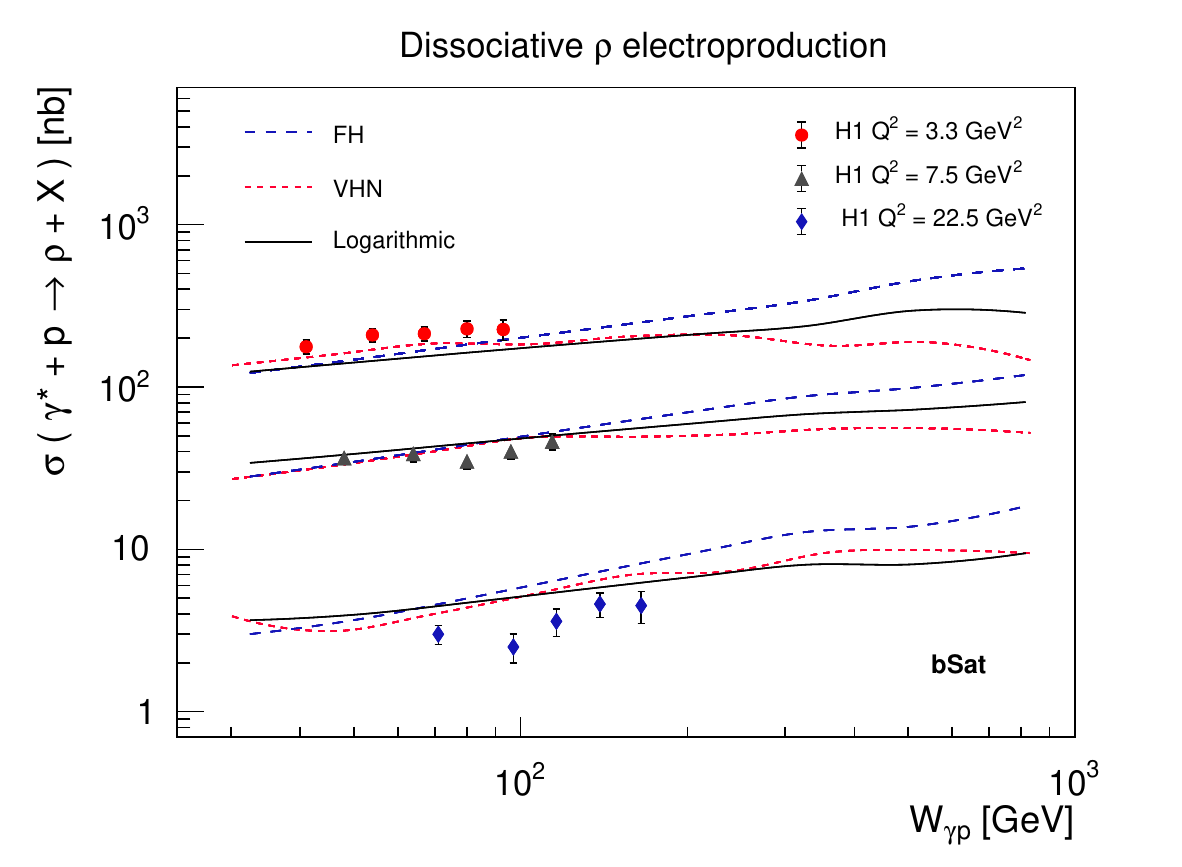}\hskip 0.5cm
	\includegraphics[width=0.45\linewidth]{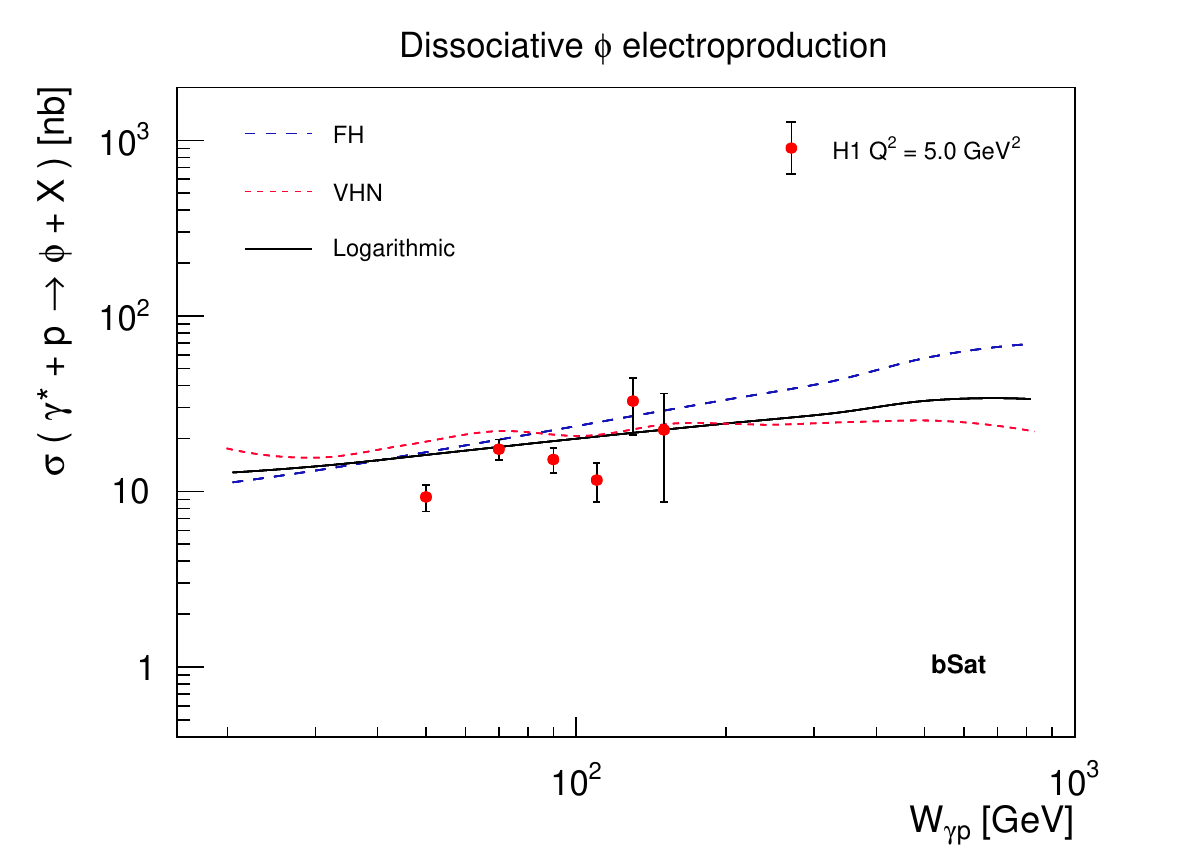}\hskip0.5cm
	\includegraphics[width=0.45\linewidth]{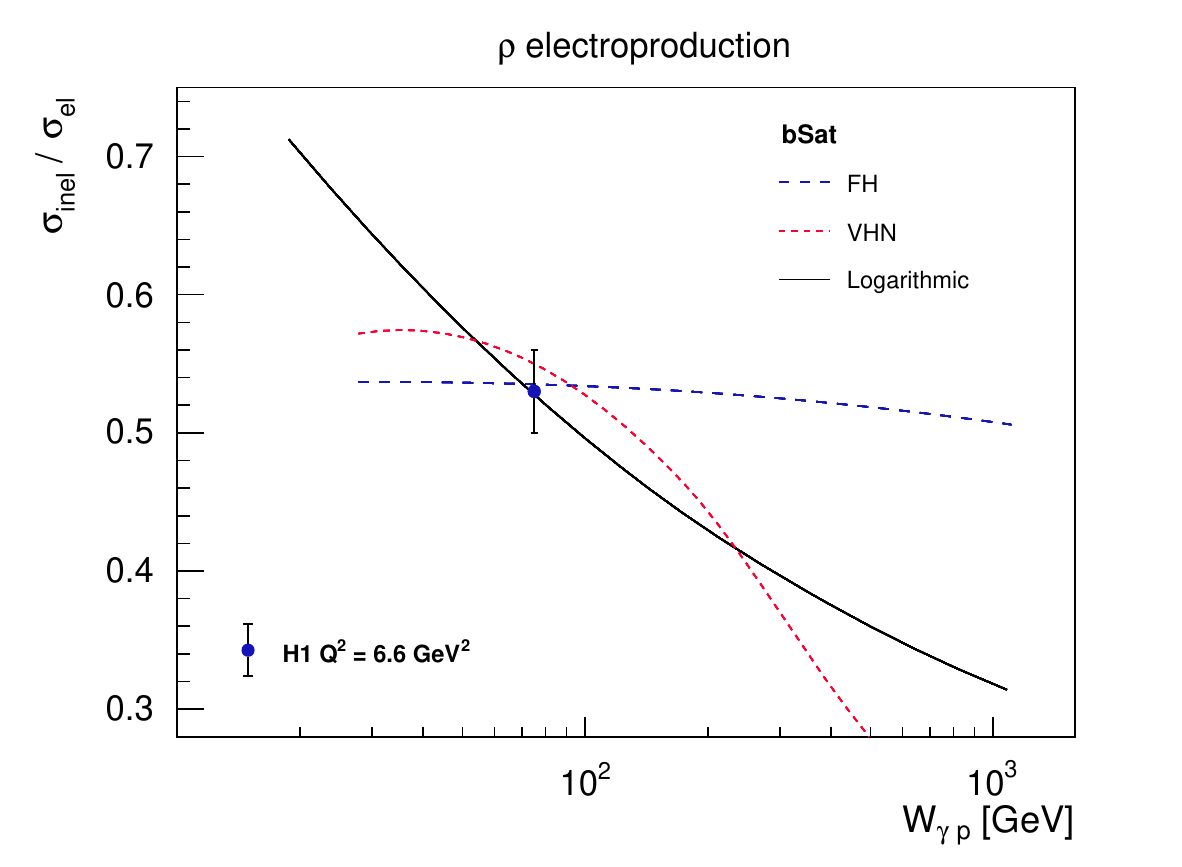}\hskip 0.5cm
	\includegraphics[width=0.45\linewidth]{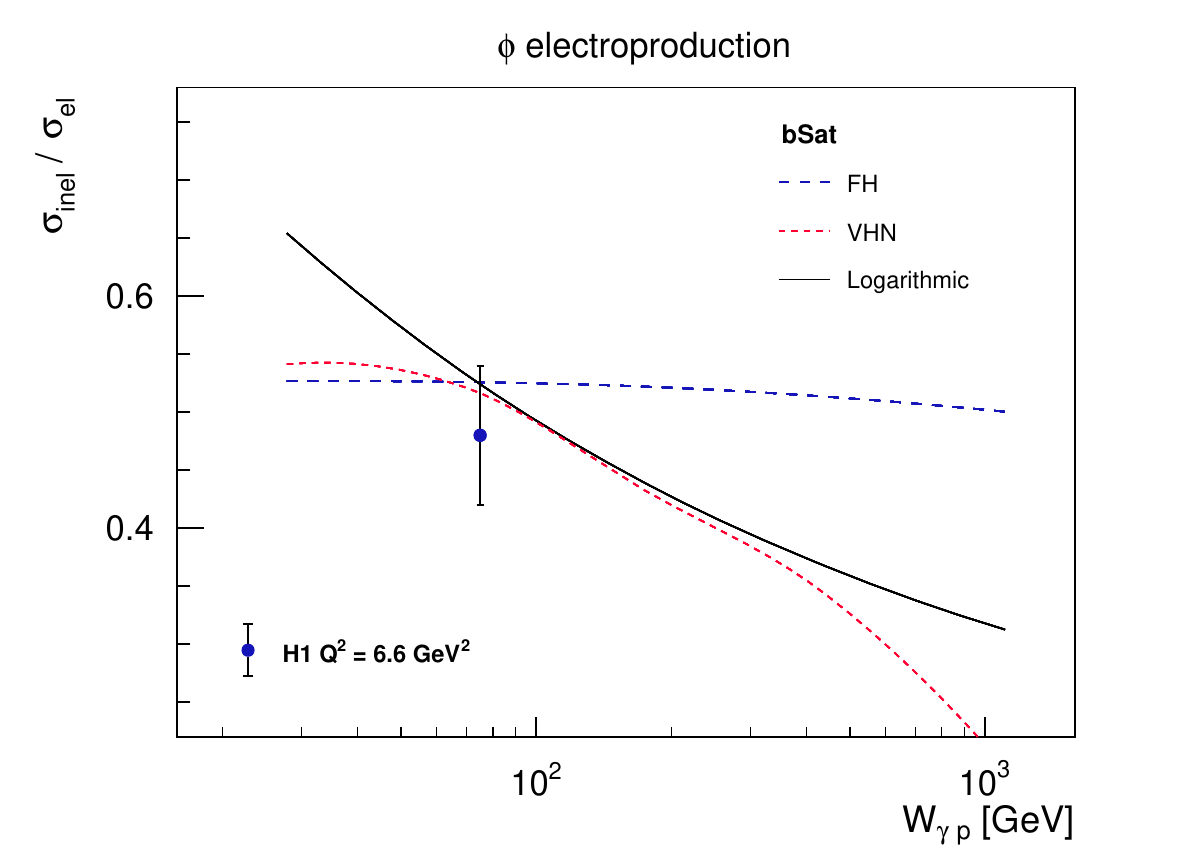}
	\caption{Energy dependence of the incoherent cross section (top) and the ratio of incoherent to coherent cross sections (bottom) for $\rho$ (left) and $\phi$ (right) electro-production in the saturated FH, VHW, VHN, and logarithmic models compared with the data from \cite{H1:2009cml}.}
	\label{t_icoh_rho}
\end{figure*}

In Fig.~\ref{t_icoh_rho}, we study the energy dependence of the incoherent cross section and the ratio of incoherent to coherent cross sections for $\rho$ and $\phi$ electro-production in the saturated FH, VHW  and VHN models. As for $J/\psi$ production, the incoherent cross sections are suppressed at high energies in the VHW and VHN models as compared to the FH model, but the available measurements are unable to distinguish between the models. The suppression of the incoherent cross section is more pronounced in the VHN model than in the other two. The ratio of incoherent to coherent cross section depicts the evolution of fluctuations with energy more clearly as the ratio decreases in the VHW and VHN models as we decrease  $x_{I\!\!P}$,  while this ratio remains constant for the FH model. This is similar to what we observed for the energy dependence of the J/$\psi$ meson. 
\section{Conclusions and Discussion}
We have considered three approaches for including energy dependence of the proton geometry into the bSat (also known as IP-Sat) and bNonSat dipole models for describing exclusive diffractive rates. Firstly, we considered the spherical proton, where we allowed the proton's gluonic width to increase with decreasing $x_{I\!\!P}$, using a simple power parametrisation where the proton width becomes $B_{G}(x_{I\!\!P})=B_{p}x_{I\!\!P}^{\lambda_{p}}$. We noted that this improved the descriptions of the  $t$-spectrum for coherent $J/\psi$ production at different $W_{\gamma p}$. However, as this model does not include any initial state fluctuations it cannot describe incoherent diffraction.

We also modified the hotspot model in three different ways. Firstly, in the VHW model, by allowing the hotspots' width to increase as $B_{q}(x_{I\!\!P})=B_{hs}x_{I\!\!P}^{\lambda_{hs}}$ and in the logarithmic model as $B_q (x_{I\!\!P}) = b_0$ln$^2(\frac{x_0}{x_{I\!\!P}})$. Thirdly, in the VHN model we allow the number of hotspots to increase at smaller $x_{I\!\!P}$ following the parametrisation of \cite{Cepila:2016uku}. We note that the VHW and logarithmic models only need one new parameter, while the VHN model needs three. We show that these approaches give similar (but distinct) suppressions of the incoherent cross section at large $W_{\gamma p}$, as the hotspots begin to overlap. However, there is a clear difference between these approaches in the slope of the $t$-spectrum in exclusive diffraction. The slope increases with $W_{\gamma p}$ in the VHW model, as well as in the logarithmic model where it grows similarly but slightly slower, while it remains constant in the VHN model. 

Currently, the available measurements are unable to distinguish between these scenarios. As seen in table \ref{table}, all models give a similar fit result. However, considering the $W_{\gamma p}$ spectrum of incoherent diffraction and the incoherent to coherent ratio, as well as the $t$-slope show a preference for the VHW and logarithmic models. This situation may improve drastically with new measurements. The EIC will be able to measure with increased precision at the small-to-medium $W_{\gamma p}$ region. These measurements would be able to resolve the tension between the H1 and ZEUS measurements which can be seen in Fig. \ref{master}, and help restrict the model parameters of the proton's profile further. The LHeC would be able to significantly extend the current phase space by probing the proton's gluonic radius and the incoherent suppression at large $W_{\gamma p}$, which will largely improve our understanding of the $x$-dependence of the proton's profile and the fluctuations spectrum. Similarly, current measurements of exclusive vector mesons in $p$A collisions in UPC at LHC and RHIC can potentially measure the incoherent to coherent ratio at large $W_{\gamma p}$ in the near future. We note that the VHW model has an exponential dependence on rapidity, which will not be physical if the phase space of the analysis is extended, which makes the logarithmic model the strongest candidate for taking this study further. 

One may also consider introducing a hybrid model in which both the hotspot width and number vary with the momentum fraction, or even vary around a mean for fixed $x_{I\!\!P}$. Conceptually, the interpretation of such a model would become ambiguous. As the  hotspots begin to overlap, the effective number of hotspots is not well defined, and the description with fewer colder hotspots becomes indistinguishable from having more and hotter hotspots. It is, therefore, phenomenologically clearer to keep the model of the proton geometry as simple as the experimental measurements allow. Thus, such a hybrid model would not be desirable.

In the future, it would be interesting to extend these models to AA ultra-peripheral collisions at the RHIC and LHC experiments, and to electron-ion collisions for the EIC, as we expect the hotspots' overlap to become more pronounced at lower $W_{\gamma p}$ due to the heavy nucleus geometry. We also plan to implement these models into the event generator sar{\emph t}re framework \cite{Toll:2013gda,Toll:2012mb}.

\subsection*{Acknowledgements}

  We are grateful to Bj\"orn Schenke, Heikki M\"antysaari and Thomas Ullrich for valuable discussions which lead to a better understanding of these phenomena and a better paper. The work of A. Kumar is supported by the Department of Science \& Technology, India under DST/INSPIRES/03/2018/000344. We thank IIT Delhi and the physics department.
 
\bibliographystyle{elsarticle-num}
\bibliography{bibliography}

    \end{document}